\documentstyle[12pt]{article}

\begin{document}

\setcounter{page}{1}

\pagestyle{plain} \vspace{1cm}
\begin{center}
\Large{\bf Bouncing universe with the non-minimally coupled quintom matter on the warped DGP Brane}\\
\small \vspace{1cm} {\bf Kourosh
Nozari$^{a,}$\footnote{knozari@umz.ac.ir}}, \quad {\bf M. R.
Setare$^{b,}$\footnote{rezakord@ipm.ir}}\quad, \quad {\bf Tahereh
Azizi$^{a,}$\footnote{t.azizi@umz.ac.ir}}\quad\\ and \\
\quad {\bf Siamak Akhshabi$^{a,}$\footnote{s.akhshabi@umz.ac.ir}} \\
\vspace{0.5cm} {\it $^{a}$Department of Physics, Faculty of Basic
Sciences,\\
University of Mazandaran,\\
P. O. Box 47416-95447, Babolsar, IRAN}\\
{\it $^{b}$ Department of Physics, Payam-e Nour University, Bijar, IRAN}\\

\end{center}
\vspace{1.5cm}
\begin{abstract}
We construct a quintom dark energy model with two non-minimally
coupled scalar fields, one quintessence and the other phantom field,
confined on the warped DGP brane. We study some important issues
such as phantom divide line crossing, existence of the bouncing
solutions and the stability of the solutions in this framework. We
show that this model accounts for crossing of the phantom divide
line and realization of the bouncing solutions. This model allows
for stability of the solutions in separate regions of the
$\omega$-$\omega'$ phase-plane.
\end{abstract}
\vspace{2cm}
\newpage
\section{Introduction}

Despite all of its successes, the standard model of cosmology
suffers from a series of problems. The most serious of these
problems is the problem of initial singularity because the laws of
physics break down at the singularity point. In order to avoid this
lawlessness, there is a huge interest in the solutions that do not
display divergencies. These solutions could be obtained at a
classical level or by quantum modifications. Most of the efforts in
quantum gravity is devoted to reveal the nature of the initial
singularity and to understand the origin of matter,
non-gravitational fields, and the very nature of the spacetime. In
recent analysis done within the loop quantum cosmology, the Big Bang
singularity is replaced by a quantum Big Bounce with finite energy
density of matter. This scenario has strong quantum effects at the
Planck scale too. Another motivation to remove the initial
singularity is the initial value problem. A sound gravitational
theory needs to have a well posed Cauchy problem. Due to the fact
that the gravitational field diverges at the singularity, we could
not have a well formulated Cauchy problem as we cannot set the
initial values at that point.

On the other hand, one of the most important discoveries over the
past few years is that we live in a positively accelerated universe
which is almost spatially flat [1]. Another remarkable hint of the
cosmological observations is that the equation of state parameter
($\omega$) transits from $\omega>-1$ to $\omega<-1$ [2,3,4]. These
discoveries generated renewed interest in bouncing models of the
universe because it can be shown that at a positively accelerated
universe a necessary condition for a bounce in general relativity is
to violate the null energy condition, i.e to have $\rho+p< 0$. To
interpret the cosmic acceleration, a so-called dark energy component
has been proposed. On the other hand the nature of dark energy is
ambiguous. The simplest candidate of dark energy is a cosmological
constant with the equation of state parameter $\omega=-1$. However,
this scenario suffers from serious problems like a huge fine tuning
and the coincidence problem [5]. Alternative models of dark energy
suggest a dynamical form of dark energy, which is often realized by
one or two scalar fields. In this respect, dark energy components
such as quintessence, k-essence, chaplygin gas, phantom and quintom
fields have been studied extensively [6] ( see also [7] and [8]).
Another alternative approach to explain the universe's late-time
acceleration is modifying the General Relativity itself [9]. Also,
some braneworld scenarios are other successful models to achieve
this goal [10]. In a braneworld scenario, our 3-brane is embedded in
a higher dimensional bulk. Matter fields are confined to a four
dimensional brane while gravity and possibly non-standard matter
fields are free to propagate in entire space time. Among the
braneworld models, the Randall-Sundrum II (RSII) model is very
popular since it has a new modification of the gravitational
potential in the very early stages of the universe evolution [11].
On the other hand, the Dvali-Gabadadze and Porrati (DGP) braneworld
scenario is a very interesting model which can describe the origin
of the late-time accelerating behavior of the universe without
adopting any additional mechanism [12]. In this setup, gravity is
modified at large distances because of an induced four-dimensional
Ricci scalar on the brane. This term can be obtained by the quantum
interaction between the matter confined on the brane and the bulk
gravitons. The DGP braneworld scenario explains accelerated
expansion of the universe via leakage of gravity to extra dimension
without need to introduce a dark energy component. While the RSII
model produces ultra-violet modification to the General Relativity,
the DGP model leads to infra-red modification. By considering the
effect of an induced gravity term as a quantum correction in RSII
model, we have a combined model that dubbed \emph{warped DGP
braneworld} in the literature [13]. This setup gives also a
self-accelerating phase in the brane cosmological evolution.

While DGP-inspired models essentially have the capability to explain
late-time acceleration, crossing of the cosmological constant line
and issues such as realization of bouncing solutions and their
stability need additional mechanism to be explained in these models.
With this viewpoint, in this paper we construct a quintom dark
energy model with two scalar fields non-minimally coupled to induced
gravity on the warped DGP brane. We study some currently important
cosmological issues such as phantom divide line crossing, avoiding
singularities by realization of the bouncing solutions and stability
of these solutions. We analyze the parameter space of the model
numerically and we show that this model allows for stability of the
solutions in the separate regions of the $\omega$-$\omega'$
phase-plane.

\section{Warped DGP Brane}
The action of the warped DGP model can be written as follows
\begin{equation}
{\cal{S}}={\cal{S}}_{bulk}+{\cal{S}}_{brane},
\end{equation}
\begin{equation}
{\cal{S}}=\int_{bulk}d^{5}X\sqrt{-{}^{(5)}g}\bigg[\frac{1}{2\kappa_{5}^{2}}
{}^{(5)}R+{}^{(5)}{\cal{L}}_{m}\bigg]+\int_{brane}d^{4}x\sqrt{-g}\bigg[\frac{1}{\kappa_{5}^{2}}
K^{\pm}+{\cal{L}}_{brane}(g_{\alpha\beta},\psi)\bigg].
\end{equation}
Here ${\cal{S}}_{bulk}$ is the action of the bulk,
${\cal{S}}_{brane}$ is the action of the brane and ${\cal{S}}$ is
the total action. $X^{A}$ with $A=0,1,2,3,5$ are coordinates in the
bulk while $x^{\mu}$ with $\mu=0,1,2,3$ are induced coordinates on
the brane. $\kappa_{5}^{2}$ is the 5-dimensional gravitational
constant. ${}^{(5)}R$ and ${}^{(5)}{\cal{L}}_{m}$ are the
$5$-dimensional Ricci scalar and the matter Lagrangian respectively.
$K^{\pm}$ is trace of the extrinsic curvature on either side of the
brane. ${\cal{L}}_{brane}(g_{\alpha\beta},\psi)$  is the effective
4-dimensional Lagrangian on the brane. The action ${\cal{S}}$ is
actually a combination of the Randall-Sundrum II and DGP model. In
other words, an induced curvature term is appeared on the brane in
the Randall-Sundrum II model, hence the name {\it Warped} DGP
Braneworld [13]. Now consider the brane Lagrangian as follows
\begin{equation}
{\cal{L}}_{brane}(g_{\alpha\beta},\psi)=\frac{\mu^2}{2}R-\lambda+L_{m},
\end{equation}
where $\mu$ is a mass parameter, $R$ is the Ricci scalar of the
brane, $\lambda$ is the tension of the brane and $L_{m}$ is the
Lagrangian of the other matters localized on the brane. We assume
that bulk contains only a cosmological constant, $^{(5)}\Lambda$.
With these choices, action (1) gives either a generalized DGP or a
generalized RS II model: it gives DGP model if $\lambda=0$ and
$^{(5)}\Lambda=0$, and gives RS II model if $\mu=0$ [13]. The
generalized Friedmann equation on the brane is as follows [13]
\begin{equation}
H^{2}+\frac{k}{a^{2}}=\frac{1}{3\mu^2}\bigg[\rho+\rho_{0}\Big(1+\varepsilon
{\cal{A}}(\rho,a)\Big)\bigg],
\end{equation}
where $\varepsilon=\pm 1$ is corresponding to two possible branches
of the solutions ( two possible embedding of the brane in the
AdS$_{5}$ bulk) in this warped DGP model and
${\cal{A}}=\bigg[{\cal{A}}_{0}^{2}+\frac{2\eta}{\rho_{0}}
\Big(\rho-\mu^{2}\frac{{\cal{E}}_{0}}{a^{4}}\Big)\bigg]^{1/2}$ where
\,\, ${\cal{A}}_{0}\equiv
\bigg[1-2\eta\frac{\mu^{2}\Lambda}{\rho_{0}}\bigg]^{1/2}$,\,\, $\eta
\equiv\frac{6m_{5}^{6}}{\rho_{0}\mu^{2}}$\,\, with $0<\eta\leq1$
\,\,and \,\,$\rho_{0}\equiv
m_{\lambda}^{4}+6\frac{m_{5}^{6}}{\mu^{2}}$. By definition,
$m_{\lambda}= \lambda^{1/4}$ and $m_{5}=k_{5}^{-2/3}$.\,
${\cal{E}}_{0}$ is an integration constant and corresponding term in
the generalized Friedmann equation is called dark radiation term. We
neglect dark radiation term in forthcoming arguments. In this case,
generalized Friedmann equation (4) attains the following form,
\begin{equation}
H^{2}+\frac{k}{a^2}=\frac{1}{3\mu^2}\bigg[\rho+\rho_{0}+\varepsilon
\rho_{0}\Big({\cal{A}}_{0}^{2}+\frac{2\eta\rho}{\rho_{0}}\Big)^{1/2}\bigg],
\end{equation}
where $\rho$ is the total energy density, including energy densities
of the scalar fields and dust matter on the brane:
\begin{equation}
\rho=\rho_\varphi+\rho_\sigma+\rho_{dm}.
\end{equation}
In what follows, we construct a quintom dark energy model on the
warped DGP brane.\\

\section{A Quintom Dark Energy Model on the Warped DGP Brane}

As a part of matter fields localized on the brane, we consider a
quitom field non-minimally coupled to induced gravity on the warped
DGP brane. The action of this non-minimally coupled quintom field is
given by
\begin{equation}
{\cal{S}}_{quint}=\int_{brane}d^{4}x\sqrt{-g}\Big[-\frac{1}{2}\xi
R(\varphi^{2}+\sigma^{2})-\frac{1}{2}\partial_{\mu}\varphi\partial^{\mu}\varphi+
\frac{1}{2}\partial_{\mu}\sigma\partial^{\mu}\sigma-V(\varphi,\sigma)\Big],
\end{equation}
where $\xi $ is a non-minimal coupling and $R$ is induced Ricci
scalar on the brane.\, $\varphi$ is a normal ( canonical or
quintessence) component while $\sigma$ is a phantom field. We have
assumed a conformal coupling of the scalar fields and induced
gravity. Variation of the action with respect to each scalar field
gives the equation of motion of that scalar field
\begin{equation}
\ddot{\varphi}+3H\dot{\varphi}+\xi R\varphi +\frac{\partial
V}{\partial\varphi}=0,
\end{equation}
and
\begin{equation}
\ddot{\sigma}+3H\dot{\sigma}-\xi R\sigma -\frac{\partial
V}{\partial\sigma}=0.
\end{equation}
The energy density and pressure of the quintom field are given by
the following relations
\begin{equation}
\rho_{quint}=\rho_\varphi+\rho_\sigma=\frac{1}{2}(\dot{\varphi}^{2}-
\dot{\sigma}^{2})+V(\varphi,\sigma)+6\xi
H(\varphi\dot{\varphi}+\sigma\dot{\sigma})+3\xi
H^{2}(\varphi^{2}+\sigma^{2})
\end{equation}
and
$$p_{quint}=p_\varphi+p_\sigma=\frac{1}{2}(\dot{\varphi}^{2}-
\dot{\sigma}^{2})-V(\varphi,\sigma)-2\xi(\varphi\ddot{\varphi}+2\varphi
H\dot{\varphi}+\dot{\varphi}^{2}+\sigma\ddot{\sigma}+2\sigma
H\dot{\sigma}+\dot{\sigma}^{2})$$
\begin{equation}
-\xi(2\dot{H}+3H^2)(\varphi^2+\sigma^2)
\end{equation}
In what follows, by comparing the modified Friedmann equation in the
warped DGP braneworld with the standard Friedmann equation, we
deduce the equation of state of the dark energy component. This is
reasonable since all observed features of dark energy are
essentially derivable in general relativity ( see [14] and
references therein). The standard Friedmann equation in four
dimensions is written as
\begin{equation}
H^2+\frac{k}{a^2}=\frac{1}{3\mu^2}(\rho_{dm}+\rho_{de}),
\end{equation}
where $\rho_{dm}$ is the dust matter density, while $\rho_{de}$ is
dark energy density. Comparing this equation with equation (5), we
deduce
\begin{equation}
\rho_{de}=\rho_{\varphi}+\rho_{\sigma}+\rho_{0}+\varepsilon\rho_0\Big(A_0^2+2\eta\frac{\rho}{\rho_0}\Big)^{\frac{1}{2}}.
\end{equation}
The conservation of the quintom field effective energy density can
be stated as
\begin{equation}
\frac{d\rho_{quint}}{dt}+3H(\rho_{quint}+p_{quint})=0
\end{equation}
Since the dust matter obeys the continuity equation and the Bianchi
identity keeps valid, total energy density satisfies the continuity
equation. In order to solve the Friedmann equation (5) we choose the
following potential
\begin{equation}
V(\varphi,\sigma)=(\zeta\varphi\sigma)^{2}+\frac{1}{2}m^{2}(\varphi^{2}-\sigma^{2}),
\end{equation}
where $\zeta$ is a dimensionless constant describing the interaction
between the scalar fields. With this potential, a possible solution
of our basic equations, (5), (9) and (10) with supplemented
equations (11) and (12) is as follows ( see [15] for a similar
argument)
\begin{equation}
\varphi=\sqrt{C_{0}}\,\cos(mt),\quad\quad
\sigma=\sqrt{C_{0}}\,\sin(mt)
\end{equation}
where $C_{0}$ is a parameter with the dimension of mass squared
describing the oscillating amplitude of the fields. For a flat
spatial geometry on the brane and setting\, $\rho_{dm}=0$, if we
consider low-energy limit where by assumption  \, $\rho_{de}\ll
\rho_{0}$, we find
\begin{equation}
\Big(\frac{\dot{a}}{a}\Big)^{2}\approx\frac{1}{3\mu^{2}}\bigg[(\rho_{\varphi}+\rho_{\sigma})(1+\frac{\varepsilon
\eta}{{\cal{A}}_{0}})+\rho_{0}(1+\varepsilon {\cal{A}}_{0})\bigg].
\end{equation}
Using (17) in (11), we find
\begin{equation}
\rho_{\varphi}+\rho_{\sigma}=\frac{\zeta^{2}C_{0}^{2}}{4}sin^{2}(2mt)+3\xi
H^{2}C_{0}.
\end{equation}
Therefore, Friedmann equation (17) can be rewritten as follows
\begin{equation}
H=\pm\bigg(\frac{\frac{\zeta^{2}C_{0}^{2}}{12\mu^{2}}\sin^{2}(2mt)
(1+\frac{\varepsilon
\eta}{{\cal{A}}_{0}})+\frac{\rho_{0}}{3\mu^{2}(1+\varepsilon
A_{0})}}{1-\frac{\xi C_{0}}{\mu^{2}(1+\frac{\varepsilon
\eta}{{\cal{A}}_{0}})}}\bigg)^{1/2}.
\end{equation}
There are four possible combinations of signs in this equation. We
use this result in our forthcoming arguments. Before proceeding
further, we note that one could choose the quantities in the square
root in such a way that lead to a imaginary Hubble parameter. We
avoid such cases in what follows. Also singularity points of $H$ are
treated in forthcoming arguments.

\subsection{Bouncing behavior of the model}
We start with a detailed examination of the necessary conditions
required for a successful bounce. During the contracting phase, the
scale factor a(t) is decreasing, i.e. $\dot{a}(t) < 0$, and in the
expanding phase we have  $\dot{a}(t) > 0$. At the bouncing point,
$\dot{a}(t) = 0$, and around this point $\ddot{a}(t)
> 0$ for a period of time [15,16]. Equivalently in the bouncing cosmology,
the Hubble parameter $H$ runs across zero from $H < 0$ to $H > 0$
and $H = 0$ at the bouncing point. A successful bounce requires that
around this point the following relation should be satisfied
\begin{equation}
\dot{H}=-4 \pi G \rho(1+\omega)>0
\end{equation}
So, at the bouncing point the scale factor reaches a non-zero
minimum value while the Hubble parameter reaches zero. By solving
the Friedmann equation (19) we plot the behavior of the scale factor
versus the cosmic time, $t$, for two branches of the solutions.
Figure $(1a)$ shows the behavior of $a(t)$ for $\varepsilon=+1$ and
figure $(1b)$ shows the case for $\varepsilon=-1$. As one can see,
in both branches of this DGP-inspired model, the scale factor
reaches a non-zero minimum and the universe switches between
expanding and contracting phases alternatively. As we have
emphasized, equation (19) has four alternative representations
corresponding to four possible combinations of the signs. If we
integrate this equation, we find
\begin{equation}
a(t)=a_{0}\exp\bigg[\pm\int{\bigg(\frac{\frac{\zeta^{2}C_{0}^{2}}{12\mu^{2}}\sin^{2}(2mt)
(1+\frac{\varepsilon
\eta}{{\cal{A}}_{0}})+\frac{\rho_{0}}{3\mu^{2}(1+\varepsilon
A_{0})}}{1-\frac{\xi C_{0}}{\mu^{2}(1+\frac{\varepsilon
\eta}{{\cal{A}}_{0}})}}\bigg)^{1/2}}\bigg].
\end{equation}
Other possible combinations of signs lead to only a shift in the
corresponding figures.
\begin{figure}[htp]
\begin{center}
\includegraphics{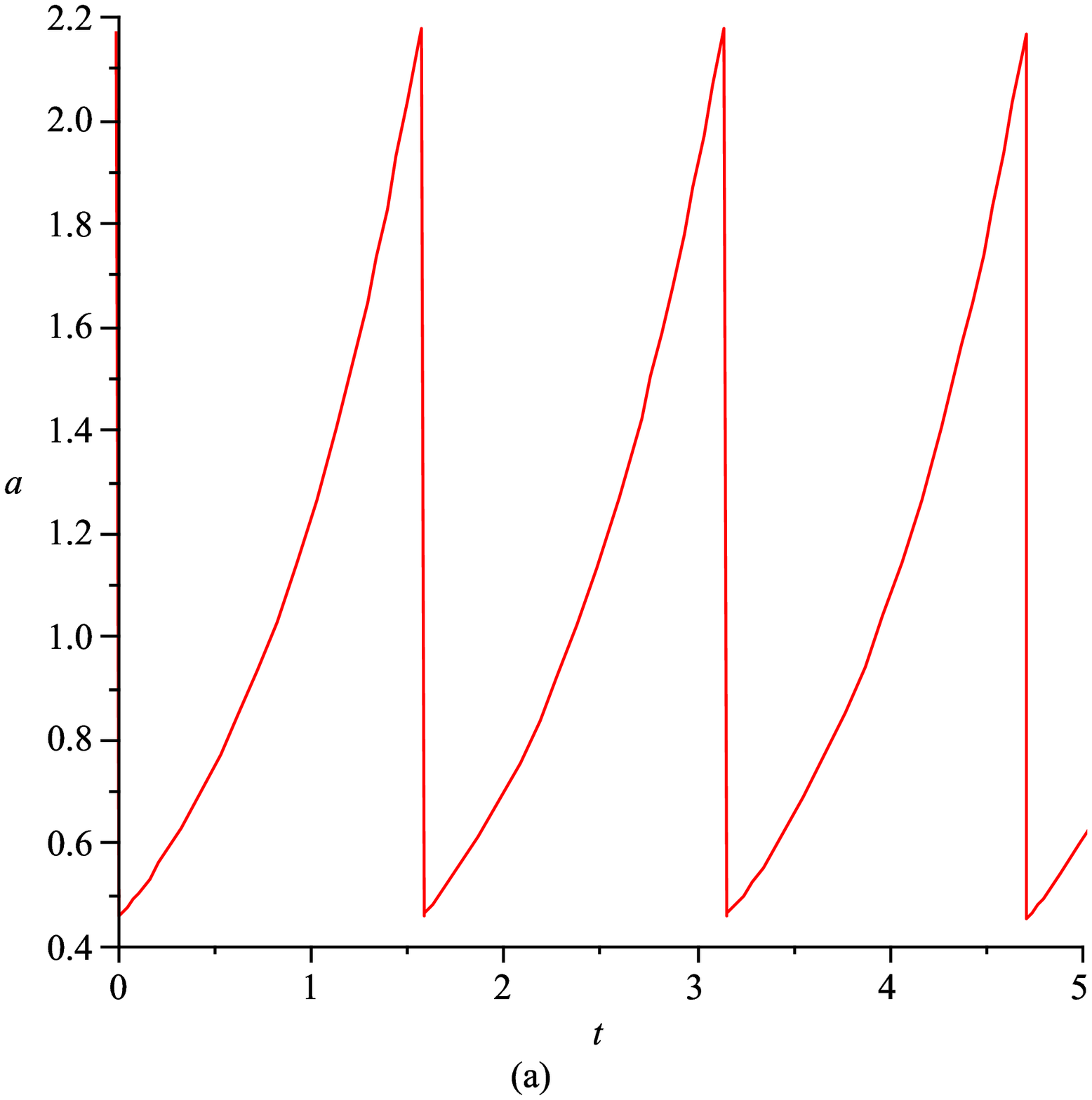} \vspace{5cm}\includegraphics{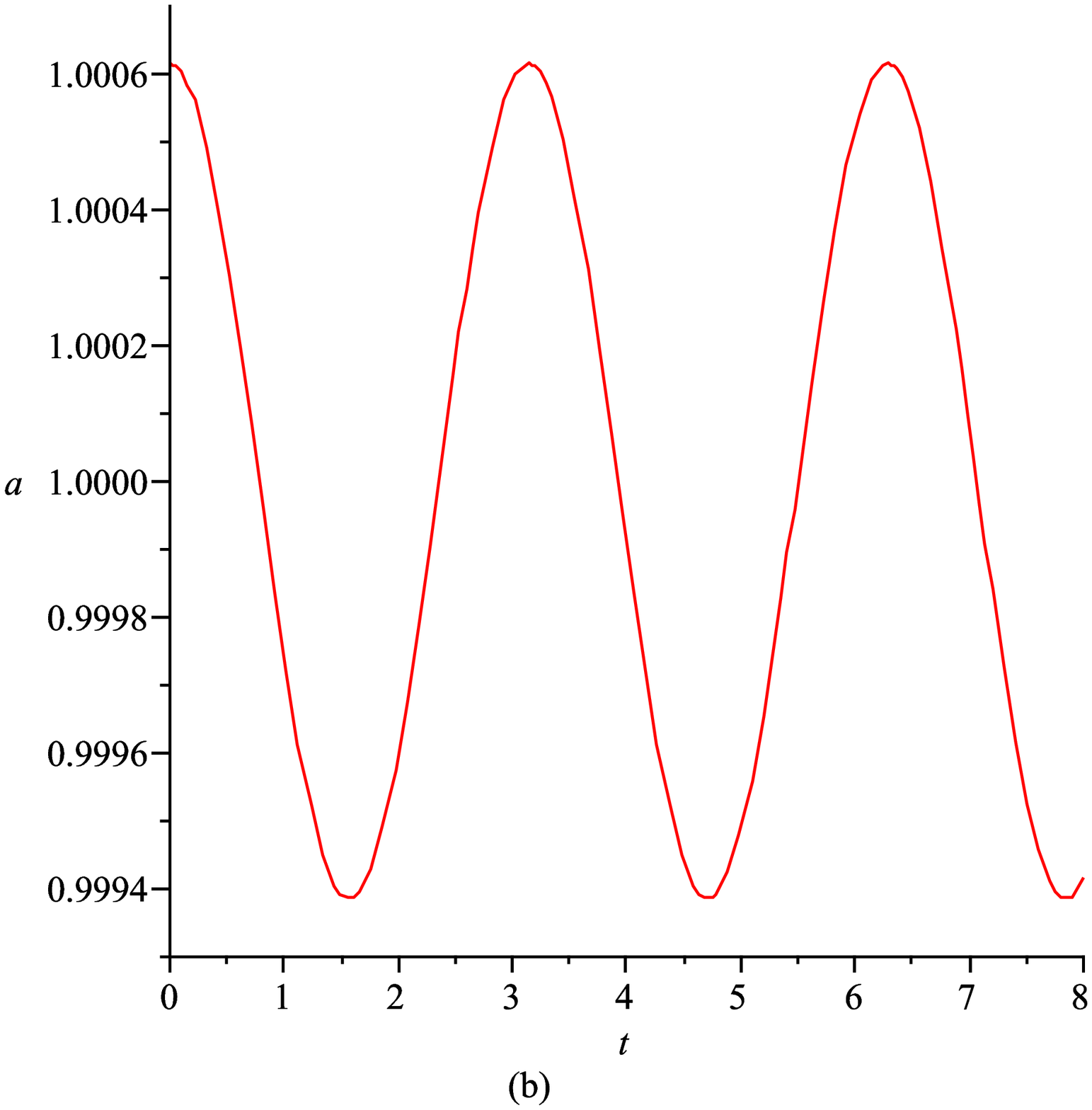}\vspace{2 cm}
\end{center}
\vspace{1cm}
 \caption{\small {The evolution of the scale factor for two branches
 of the warped DGP model with quintom field localized on the brane:
 a) Self-accelerating branch of the model ( with $\varepsilon=+1$),
 the universe undergoes an expansion, reaches to a maximum radius and then
 crunches to a finite minimum size and this cycle repeats.
 There is no bounce at the minimum point since the scale
 factor has no derivative at that point. b) Normal branch of the model
 ( with $\varepsilon=-1$). The universe switches alternatively between
 expanding and contracting phases. The minimum points are the bouncing points. }}
\end{figure}
\newpage
\subsection{Crossing of the phantom divide line}
In the DGP scenario if we use a single scalar field (ordinary or
phantom ) on the brane, we can show that the equation of state
parameter of dark energy crosses the phantom divide line [17] ( see
also [14] and [18]). It has been shown that DGP model with a quintom
dark energy fluid in the bulk or brane, accounts for the phantom
divide line crossing too [19]. Now we try to realize this crossing
in the \emph{warped} DGP braneworld with quintom matter localized on
the brane and non-minimally coupled to induced gravity. In this
warped DGP model, the equation of state parameter, $\omega$ of dark
energy component has the following form ( with $\rho_{dm}=0$)
$$\omega=-1+\frac{(\dot{\varphi}^2-\dot{\sigma}^2)-2\xi\Big[-H(\varphi\dot{\varphi}+
\sigma\dot{\sigma})
+\dot{H}(\varphi^2+\sigma^2)+\varphi\ddot{\varphi}+\sigma\ddot{\sigma}+\dot{\varphi}^2+
\dot{\sigma}^2\Big]}{\rho_{de}}\times
$$
\begin{equation}
\Bigg\{1+\varepsilon\eta\bigg(A_{0}^2+2\eta\frac{\frac{1}{2}(\dot{\varphi}^{2}-\dot{\sigma}^{2})+V(\varphi,\sigma)+6\xi
H(\varphi\dot{\varphi}+\sigma\dot{\sigma})+3\xi
H^{2}(\varphi^{2}+\sigma^{2}))}{\rho_0}\bigg)^{-\frac{1}{2}}\Bigg\}
\end{equation}
After substituting corresponding relations for $\varphi$, $\sigma$,
$H$ and $V$ in this equation, we plot the behavior of $\omega$ for
two branches of the DGP-inspired model versus the cosmic time.
Figure $2$ shows the variation of $\omega$ versus cosmic time for
two possible branches of the model. In figure $2a$ which is devoted
to self-accelerating branch, the equation of state parameter crosses
the cosmological constant line. This behavior is repeated
periodically due to oscillating nature of the cosmic expansion.
Figure $2b$ shows the situation for normal ( non-self accelerating)
branch. In this case crossing of the cosmological constant line
occurs too, but the behavior of equation of state parameter differs
considerably compared to self-accelerating branch. As this figure
shows, at the bouncing point $\omega$ approaches the negative
infinity.
\begin{figure}[htp]
\begin{center}
\includegraphics{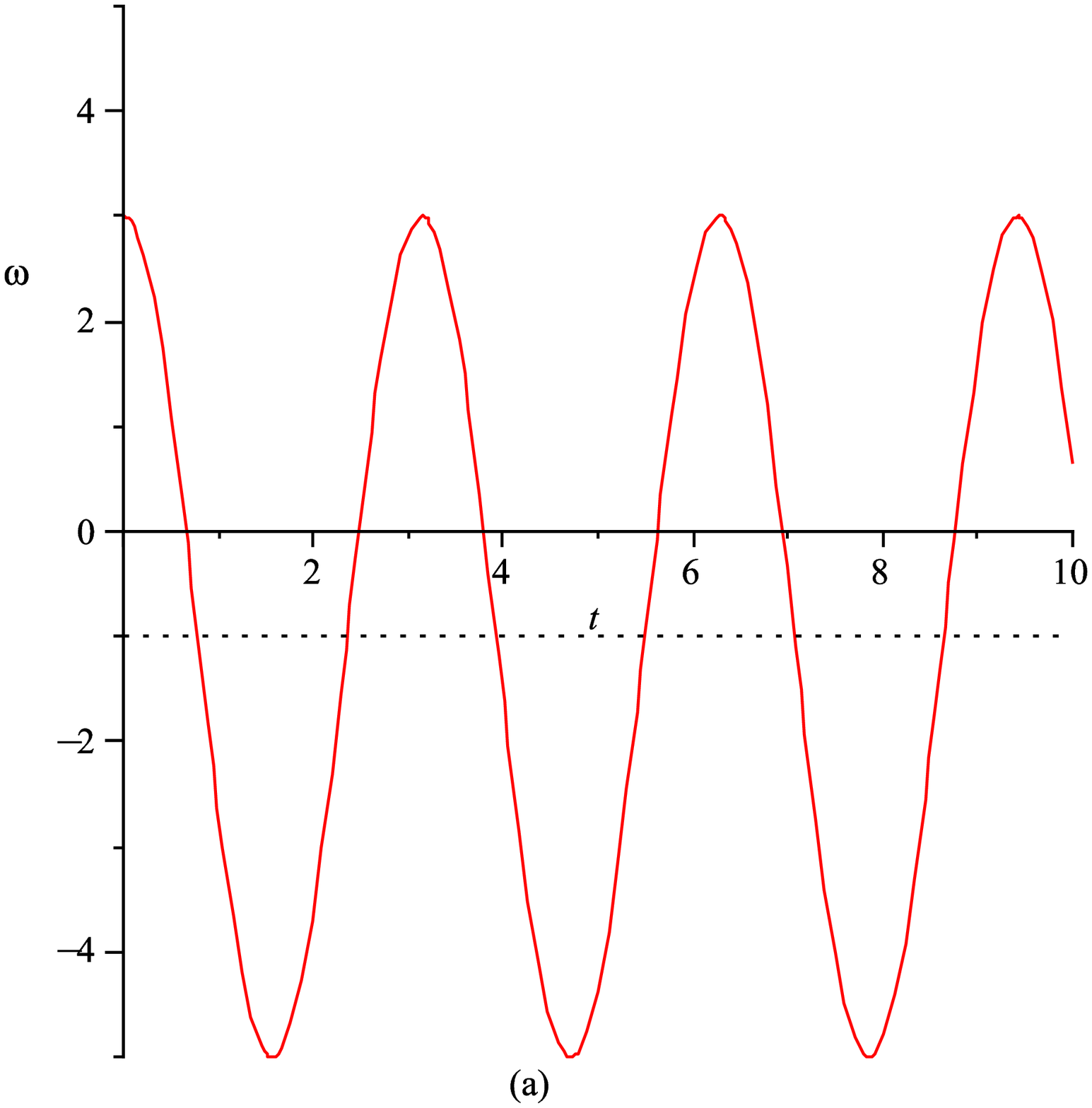} \vspace{5cm}\includegraphics{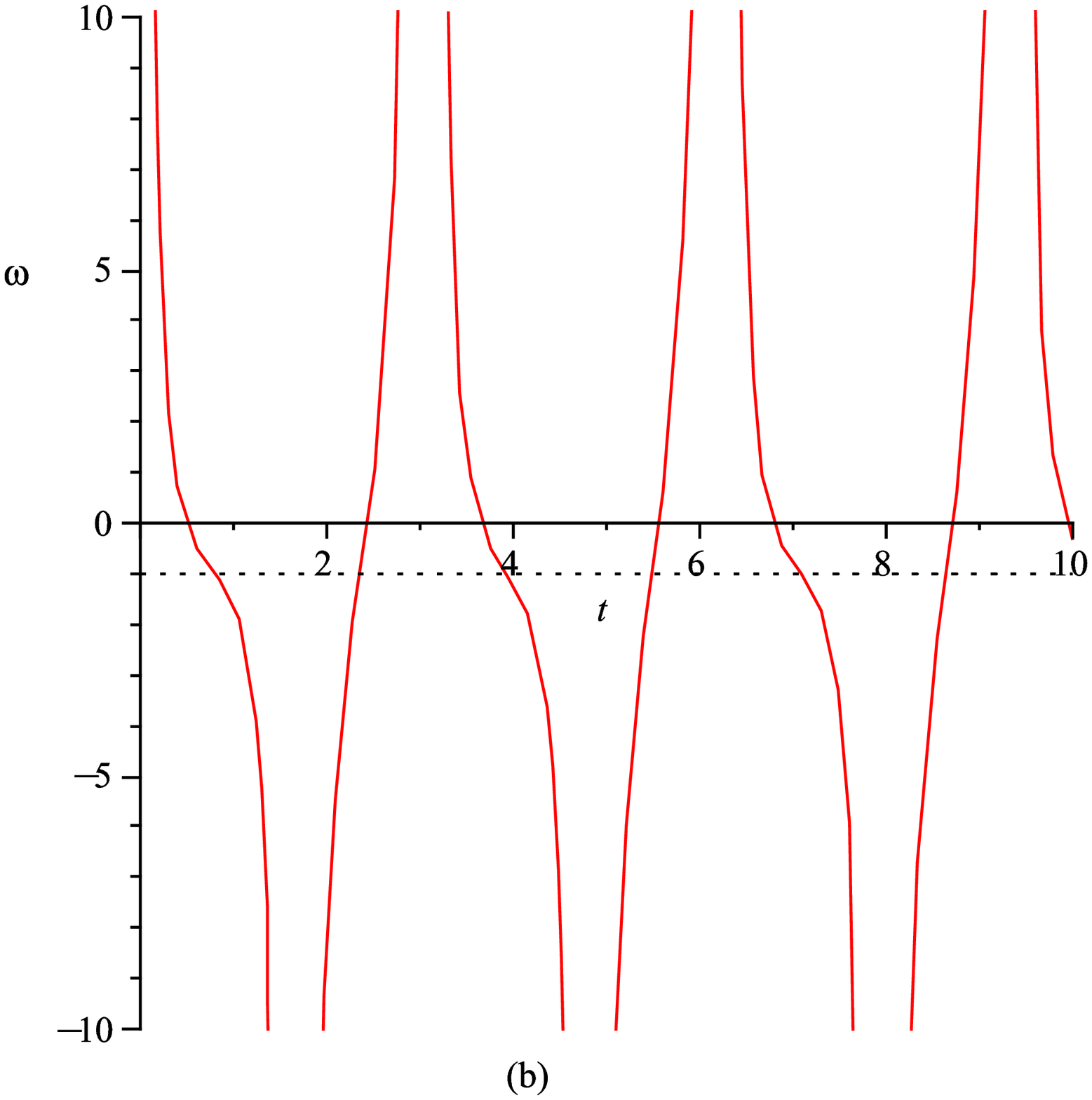}\vspace{2cm}
\end{center}
\vspace{2.5cm}
 \caption{\small {Time evolution of the equation of state parameter $\omega$.
 There is a crossing of the cosmological constant line in both branches of the scenario:
  a) Self-accelerating branch. $\omega$ mimics the oscillating nature of the cosmic expansion.
   b) Normal branch. As usual, at the bouncing
point $\omega$ approaches the negative infinity.}}
\end{figure}
\newpage
\subsection{Stability of the model}
Now we study the stability of our model. The sound speed expresses
the phase velocity of the inhomogeneous perturbations of the quintom
field. In order to study the classical stability of our model, we
analyze the behavior of the model in the $\omega-\omega'$ plane
where $\omega'$ is the derivative of $\omega$ with respect to the
logarithm of the scale factor ( see [20-23] for a similar analysis
for other interesting cases)
\begin{equation}
\omega'\equiv\frac{d\omega}{d\ln a}=\frac{d\omega}{dt}\frac{dt}{d\ln
a}=\frac{\dot{\omega}}{H}.
\end{equation}
We define the function $c_{a}$ as
\begin{equation}
c^{2}_{a}\equiv\frac{\dot{p}}{\dot{\rho}}\,.
\end{equation}
If the matter is considered as a perfect fluid, this function would
be the adiabatic sound speed of this fluid. But, for our model with
two scalar fields, this is not actually a sound speed. Nevertheless,
we demand that $c^{2}_{a}>0$ in order to avoid the big rip
singularity at the end of the universe evolution. From equation (14)
we have
\begin{equation}
\dot{\rho}_{de}=-3H\rho_{de}(1+\omega_{de}).
\end{equation}
Using equation of state $p_{de}=\omega_{de}\rho_{de}$, we get
\begin{equation}
\dot{p}_{de}=\dot{\omega}_{de}\rho_{de}+\omega_{de}\dot{\rho}_{de}.
\end{equation}
So, the function $c^{2}_{a}$ could be rewritten as
\begin{equation}
c^{2}_{a}=\frac{\dot{\omega}_{de}}{-3H(1+\omega_{de})}+\omega_{de}.
\end{equation}
In this situation, the $\omega-\omega'$ plane is divided into four
regions defined as follows
\begin{equation}
 \left\{\begin{array}{c}
   I:   \quad \,\omega_{de}>-1,\quad\quad \omega'>3\omega(1+\omega) \quad \Rightarrow \quad\quad c^{2}_{a}>0 \\
   II:  \quad\,\omega_{de}>-1,\quad\quad \omega'<3\omega(1+\omega)\quad \Rightarrow \quad\quad c^{2}_{a}<0 \\
   III: \quad\,\omega_{de}<-1, \quad\quad  \omega'>3\omega(1+\omega)\quad \Rightarrow \quad\quad c^{2}_{a}<0 \\
   IV:  \quad\,\omega_{de}<-1,\quad\quad \omega'<3\omega(1+\omega)\quad \Rightarrow\quad\quad c^{2}_{a}>0 \\
  \end{array}\right.
\end{equation}
As one can see from these relations, the regions I and IV have the
classical stability in our model. We plot the behavior of the model
in the $\omega-\omega'$ phase plane and identify the regions
mentioned above in figure $3$.
\begin{figure}[htp]
\begin{center}
\includegraphics{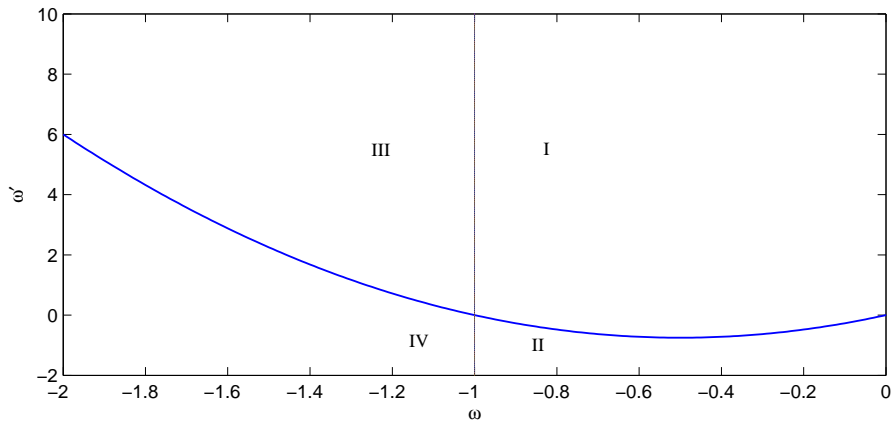}
\end{center}
\vspace{4 cm}
 \caption{\small { Bounds on $\omega'$ as a function of $\omega$
 in $\omega-\omega'$ phase plane.
 The stable regions are I and IV.}}
\end{figure}

\section{Summary}
One of the most serious shortcomings of the standard model of
cosmology is the problem of initial ( and possibly final)
singularity. In recent analysis done within the loop quantum
cosmology, the Big Bang singularity is replaced by a quantum Big
Bounce with finite energy density of matter. Also incorporation of
the Gauss-Bonnet term in the action of braneworld models with
induced gravity provides a phenomenologically rich framework to
overcome initial singularity with possible realization of bouncing
solutions [24]. On the other hand, a sound gravitational theory
needs also to have a well posed Cauchy problem. A Model universe
which realizes bouncing solution is a good candidate to overcome
these singularities.

An alternative approach to explain late-time positively accelerated
expansion of the universe is a multi-component dark energy with at
least one non-canonical phantom field. The analysis of the
properties of dark energy from recent observations mildly favor
models where $\omega=\frac{p}{\rho}$ crosses the phantom divide
line, $\omega=-1$ in the near past. In this respect, construction of
theoretical frameworks with potential to describe this positively
accelerated expansion and crossing of the phantom divide line by the
equation of state parameter is an interesting challenge. In this
paper, we have considered a quintom field non-minimally coupled to
induced gravity on the warped DGP braneworld as a dark energy
component. We have studied the bouncing behavior of the solutions in
both branches of this DGP-inspired scenario. In the
self-accelerating branch of the model ( with $\varepsilon=+1$), the
universe undergoes an expansion, reaches to a maximum radius and
then crunches to a finite minimum size and this cycle repeats. In
this case there is no bounce at the minimum point since the scale
factor has no derivative at that point. In the normal ( non-self
accelerating ) branch of the model ( with $\varepsilon=-1$), the
universe switches alternatively between expanding and contracting
phases. The minimum points of the scale factor versus cosmic time
are the bouncing points. In fact there is a sequence of phases as:
Expansion $\rightarrow$ Turn-around $\rightarrow$ Contraction
$\rightarrow$ Bounce and this cycle repeats regularly.

Next we study the dynamics of the equation of state parameter. One
can see that there is a crossing of the phantom divide line in both
branches of this DGP-inspired model although the evolution of the
equation of state parameter is different in these two branches. We
have studied the stability of this model. As a result, there are
appropriate regions of $\omega-\omega'$ phase plane that solutions
are stable.

Finally we should stress on the ghost instabilities present in the
self-accelerating branch of this DGP-inspired model. The
self-accelerating branch of the DGP model contains a ghost at the
linearized level [25]. Since the ghost carries negative energy
density, it leads to the instability of the spacetime. The presence
of the ghost can be attributed to the infinite volume of the
extra-dimension in DGP setup. When there are ghosts instabilities in
self-accelerating branch, it is natural to ask what are the results
of solutions decay. As a possible answer we can state that since the
normal branch solutions are ghost-free, one can think that the
self-accelerating solutions may decay into the normal branch
solutions. In fact for a given brane tension, the Hubble parameter
in the self-accelerating universe is larger than that of the normal
branch solutions. Then it is possible to have nucleation of bubbles
of the normal branch in the environment of the self-accelerating
branch solution. This is similar to the false vacuum decay in de
Sitter space. However, there are arguments against this kind of
reasoning which suggest that the self-accelerating branch does not
decay into the normal branch by forming normal branch bubbles ( see
[25] for more details). It was also shown that the introduction of
Gauss-Bonnet term in the bulk does not help to overcome this problem
[26]. In fact, it is still unclear what is the end state of the
ghost instability in self-accelerated branch of DGP inspired setups.
On the other hand, non-minimal coupling of scalar field and induced
gravity in our setup provides a new degree of freedom which requires
special fine tuning and this my provide a suitable basis to treat
ghost instability. It seems that in our model this additional degree
of freedom has the capability to provide the background for a more
reliable solution to ghost instability due to wider parameter space.

\end{document}